\newcommand{\gray}{\mbox{$\gamma$-ray}}
\newcommand{\pcmq}{\mbox{cm$^{-2}$}}

\newcommand{\psec}{\mbox{s$^{-1}$}}
\newcommand{\funit}{\mbox{ph \pcmq \psec}}

\def\deg{\ensuremath{^\circ}}

\documentclass{aa}
\usepackage{epsfig}
\begin{document}

\title{Early SPI/INTEGRAL constraints on the morphology of the 511~keV 
line emission in the 4th galactic quadrant\thanks{Based on observations with
INTEGRAL, an ESA project with instruments and science data centre
funded by ESA member states (especially the PI countries: Denmark,
France, Germany, Italy, Switzerland, Spain), Czech Republic and
Poland, and with the participation of Russia and the USA.}}

\author{J.~Kn\"odlseder\inst{1}
         \and V.~Lonjou\inst{1}
         \and P.~Jean\inst{1}
         \and M.~Allain\inst{1}
	 \and P.~Mandrou\inst{1}
         \and J.-P.~Roques\inst{1}
         \and G.~K.~Skinner\inst{1}
         \and G.~Vedrenne\inst{1}
         \and P.~von~Ballmoos\inst{1}
	 \and G.~Weidenspointner\inst{1,5}
	 \and P.~Caraveo\inst{4}
	 \and B.~Cordier\inst{3}
         \and V.~Sch\"onfelder\inst{2}
	 \and B.~J.~Teegarden\inst{5}
}

\offprints{J\"urgen Kn\"odlseder, e-mail : knodlseder@cesr.fr}

\institute{
$^{1}$ Centre d'Etude Spatiale des Rayonnements, CNRS/UPS, B.P.~4346, 
       31028 Toulouse Cedex 4, France \\
$^{2}$ Max-Planck-Institut f\"ur Extraterrestrische Physic, Postfach 1603, 
       85740 Garching, Germany  \\
$^{3}$ DSM/DAPNIA/SAp, CEA-Saclay, 91191 Gif-sur-Yvette, France \\
$^{4}$ IASF, via Bassini 15, 20133 Milano, Italy \\
$^{5}$ Laboratory for High Energy Astrophysics, NASA/Goddard Space Flight Center, 
       Greenbelt, MD 20771, USA
}

\date{Received / Accepted }

\authorrunning{J.~Kn\"odlseder et al.}

\titlerunning{SPI/INTEGRAL constraints on the morphology of the galactic 
511 keV line emission}

\abstract{
We provide first constraints on the morphology of the 511 keV line 
emission from the galactic centre region on basis of data taken with 
the spectrometer SPI on the INTEGRAL gamma-ray observatory.
The data suggest an azimuthally symmetric galactic bulge component with 
FWHM of $\sim9\deg$ with a $2\sigma$ uncertainty range covering
$6\deg-18\deg$.
The 511 keV line flux in the bulge component amounts to
$9.9^{+4.7}_{-2.1} \times 10^{-4}$ \funit.
No evidence for a galactic disk component has been found so far;
upper $2\sigma$ flux limits in the range $(1.4-3.4) \times 10^{-3}$ 
\funit\ have been obtained that depend on the assumed disk morphology.
These limits correspond to lower limits on the bulge-to-disk ratio 
of $0.3-0.6$.
}

\maketitle

\section{Introduction}
\label{sec:intro}

Since the first detection (Johnson \& Haymes \cite{johnson1973}) and 
the subsequent identification (Leventhal et al. \cite{leventhal1978})
of the galactic 511 keV annihilation line, the origin of galactic 
positrons has become a lively topic of scientific debate.
Among the proposed source candidates figure
compact objects such as neutron stars or black holes
(Lingenfelter \& Ramaty \cite{lingenfelter1983}),
stars, such as supernovae, novae, red giants and Wolf-Rayet stars,
expelling radioactive nuclei produced by nucleosynthesis
(Ramaty, Kozlovsky \& Lingenfelter \cite{ramaty1979}),
cosmic-ray interactions with the interstellar medium
(Kozlovsky, Lingenfelter \& Ramaty \cite{kozlovsky1987}),
pulsars (Sturrock \cite{sturrock1971}),
gamma-ray bursts (Lingenfelter \& Hueter \cite{lingenfelter1984})
and stellar flares.
Yet so far the source of the galactic positrons is still unknown.

The question of the morphology of the galactic 511 keV annihilation 
signal is intimately related to the question of the origin of 
galactic positrons.
The celestial distributions should be tied to the source 
distribution, although positron diffusion within the Galaxy may to 
some extent blur this link.
Although earlier measurements already provided first indications of 
the morphology of the emission (e.g. Share et al. \cite{share1990}),
it is only with the advent of the OSSE telescope onboard the Compton 
Gamma-Ray Observatory that a first crude skymap of the 511 keV 
intensity distribution became available 
(Cheng et al. \cite{cheng1997}; Purcell et al. \cite{purcell1997};
Milne et al. \cite{milne2000}; Milne et al. \cite{milne2001}).
The OSSE observations suggest at least two emission components, one 
being a spheroidal bulge and the other being a galactic disk component.
Indications of a third component situated above the galactic plane 
have resulted in various speculations about the underlying source (von 
Ballmoos et al. \cite{ballmoos2003}), yet the morphology and intensity 
of this component is only poorly determined 
(e.g. Milne et al. \cite{milne2001}).

Modelling the bulge component by symmetric gaussians, the OSSE data 
suggest FWHM values in the range $4\deg-6\deg$ with no significant 
offset from the galactic centre
(Purcell et al. \cite{purcell1997}; Milne et al. \cite{milne2000}).
TGRS results are also consistent with no significant offset from the 
centre ($\la2\deg$) but suggest a somewhat broader distribution
(Harris et al. \cite{harris1998}).
The estimated bulge-to-disk ratio (hereafter B/D ratio) is only 
poorly constrained by the observations, and estimates vary from 
$0.2-3.3$ depending upon whether the bulge component features a halo 
(which leads to a large B/D ratio) or not.
The total flux has been constrained to $(2.1-3.1) \times 10^{-3}$ \funit\
(Milne et al. \cite{milne2000}).

Clearly, more observations are needed to better constrain the 511 keV 
emission morphology, and thus to shed light on the origin of galactic 
positrons.
In this paper we report a step towards this direction by exploiting a 
first set of data recorded by the spectrometer SPI on ESA's INTEGRAL 
observatory.
First results on the spectral shape of the 511 keV line obtained from the 
same set of data, indicating a slightly broadened line of 
$2.95^{+0.45}_{-0.51}$~keV FWHM, have been reported elsewhere 
(Jean et al. \cite{jean2003}).

\section{Data analysis}
\label{sec:analysis}

The data analysed in this work were accumulated during the first
year's Galactic Centre Deep Exposure (GCDE) and Galactic Plane Scan (GPS), 
executed as part of INTEGRAL's guaranteed time observations 
(see Winkler \cite{winkler2001}). 
We used data from 19 orbits from March 3rd to April 30th, 2003, amounting 
to a total effective exposure time of 1667 ks. 
The GCDE consists of rectangular pointing grids covering galactic longitudes 
$l=\pm30\deg$ and latitudes $b=\pm10\deg$, with reduced exposure up to
$b=\pm20\deg$. 
The GPS consists of pointings within the band $b=\pm6.4\deg$ along the 
galactic plane. 
For details see Winkler (\cite{winkler2001}). 
The present data are from 1199 pointings with an average exposure of 1400 
seconds per pointing.

\begin{figure}[t!]
   \includegraphics[width=8.8cm]{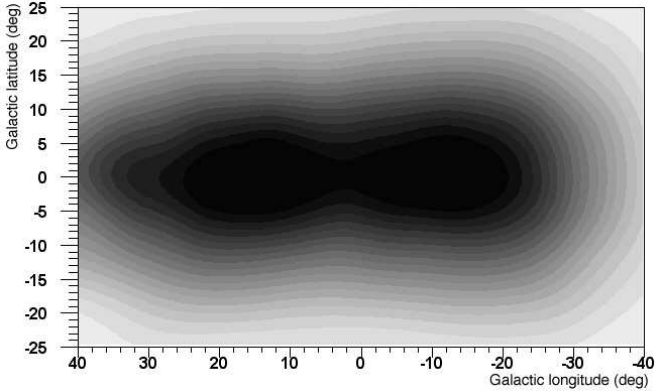}
   \caption{Relative exposure map of the galactic centre region.
   Black corresponds to regions of maximum exposure.}
    \label{fig:exposure}
\end{figure}

As a result of data sharing agreements, the results presented here are
limited to the galactic quadrant $l=270\deg$ -- $360\deg$ but, in accordance
with those agreements, data from pointings in the entire GCDE region
$l=\pm30\deg$ have been taken into account in the analysis.
The resulting sky exposure is depicted in 
Fig.~\ref{fig:exposure}, where black/white corresponds to regions of 
maximum/minimum exposure, respectively.
A quite homogeneous exposure has been achieved over galactic 
longitudes $\pm25\deg$, with a small exposure dip near the galactic 
centre.
The latitude dependence is approximated by a gaussian centred on the 
galactic plane, with FWHM of $\sim30\deg$.

Data preparation and modelling of the instrumental background is 
identical to that described in Jean et al. (\cite{jean2003}).
The SPI single detector event data have been gain corrected and binned
into event spectra of $0.25$ keV bin width for each detector and pointing, 
leading to a 3-dimensional data space.
The background has been modelled in this data space by two components,
accounting for the instrumental 511 keV line and the underlying continuum 
(see Eqs.~1 and 2 and Fig.~1 in Jean et al.~\cite{jean2003}).
The short-term variability ($<3$ days) of the background has been predicted 
using the rate of saturating events in the Ge detectors, while the 
long-term variability has been regarded as an unknown (see section 
\ref{sec:fitting}).

The expected number of counts in this data space, $E_{p,d,e}$, where 
$p$ indicates the pointing, $d$ the detector, and $e$ the energy bin,
is given by
\begin{equation}
E_{p,d,e} = \sum_{l,b} R_{p,d,e}^{l,b} \Phi_{l,b} + B_{p,d,e} ,
\end{equation}
where $R_{p,d,e}^{l,b}$ is the instrumental response matrix (i.e.~the 
number of counts in a data space bin that arises from a source of 
unit flux at the sky position $l,b$; see Sturner et al., this 
volume), $\Phi_{l,b}$ is the 511 keV sky intensity distribution, and 
$B_{p,d,e}$ is the background model.
In the following we describe two approaches that we used to extract information 
about $\Phi_{l,b}$ from the measured SPI data.

\section{Imaging}
\label{sec:imaging}

A qualitative impression about the spatial distribution of the 511~keV 
\gray\ line emission can be obtained by deconvolving the data into 
a celestial intensity distribution.
For this purpose, we employed in this work the Richardson-Lucy algorithm
(Richardson \cite{richardson1972}, Lucy \cite{lucy1974}) that has 
been successfully applied to \gray\ data of earlier missions 
(e.g. Kn\"odlseder et al. \cite{knoedlseder1999}; Milne et al. 
\cite{milne2000}).
The Richardson-Lucy algorithm decomposes the sky into a grid of 
equally sized pixels (here of $0.5\deg \times 0.5\deg$) and solves 
for the intensity $\Phi_{l,b}$ in each of the pixels simultaneously using an 
iterative maximum likelihood scheme.
The pixels are constrained to positive intensities.
As all pixel-based deconvolution algorithms, the Richardson-Lucy scheme 
leads to morphology artefacts in the case of reconstructing a diffuse 
low significance signal since the number of free parameters (i.e. the 
number of image pixels) greatly exceeds the information available in the 
data.
In order to reduce these artefacts, we added a smoothing step to the 
iterative scheme that effectively combines adjacent pixels and thus 
reduces the number of pixels in the reconstruction.
As smoothing kernel, a boxcar average of $6\deg \times 6\deg$ has been 
employed.

\begin{figure}[t!]
   \includegraphics[width=8.8cm, height=8.0cm]{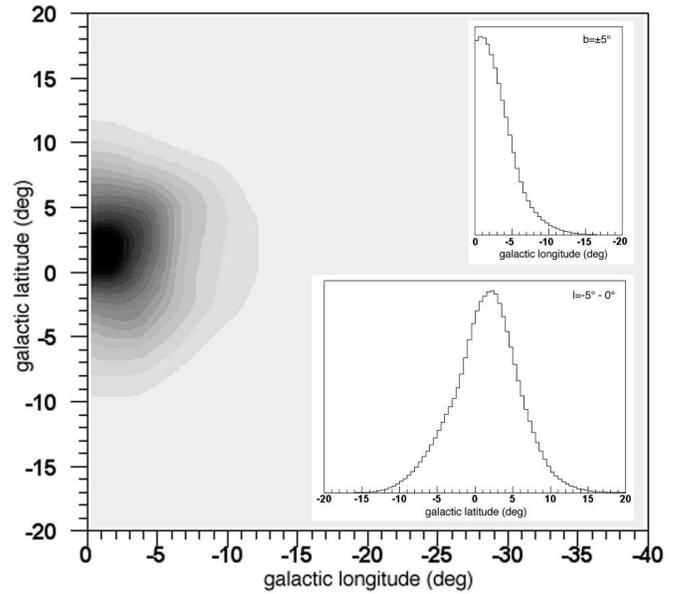}
   \caption{511~keV gamma-ray line intensity map of the galactic centre region 
   (only negative longitudes). Black corresponds to regions of maximum 
   511~keV line intensity. Longitude and latitude profiles, integrated 
   over $b=\pm5\deg$ and $l=-5\deg - 0\deg$, respectively, are shown 
   as insets.}
    \label{fig:skymap}
\end{figure}

Figure \ref{fig:skymap} shows the resulting skymap for negative 
longitudes (according to the above cited data right agreements we 
deliberately limit the image to $l<0\deg$).
As insets, longitude and latitude profiles of the 511 keV line emission
are shown that have been obtained by integrating the intensity of latitudes 
$b=\pm5deg$ and longitude $l=-5\deg - 0\deg$, respectively.
The emission is concentrated in an azimuthally symmetric region 
near the galactic centre, with an extent of about $\sim9\deg$ (FWHM). 
The emission maximum is slightly offset from the galactic centre, and 
is situated at about $l=-1\deg$ and $b=2\deg$.
Integrating the intensity over the feature suggests a flux of the 
order of $10^{-3}$ \funit.

To asses the significance of the morphology that is seen 
in the skymap, we performed Monte-Carlo simulations of the 
deconvolution process assuming that the 511 keV line emission 
morphology is described by an azimuthally symmetric gaussian bulge of 
$10\deg$ FWHM centred at $l=0\deg$ and $b=0\deg$.
In the ideal case the image deconvolution should reproduce the input 
image of an azimuthally symmetric emission centred at the galactic 
centre, yet the limited statistics may introduce some uncertainties in 
the reconstruction procedure.

\begin{figure}[t!]
   \includegraphics[width=8.8cm, height=8.0cm]{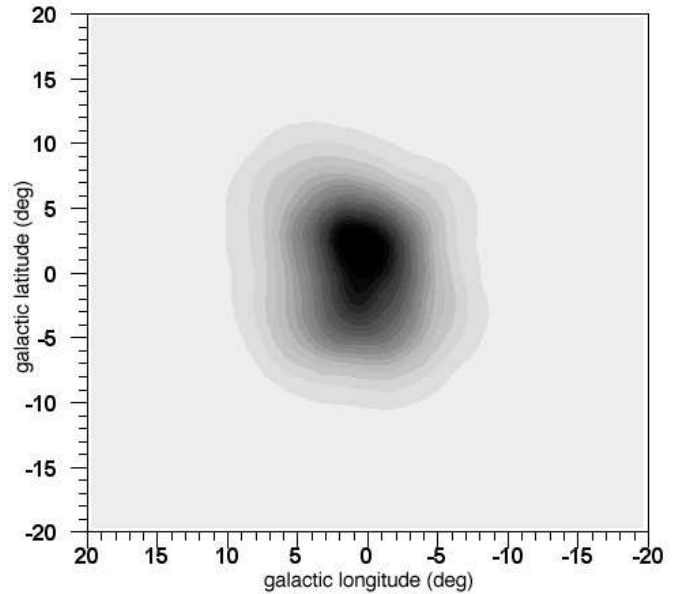}
   \caption{Simulated 511~keV gamma-ray line intensity map based on an 
   azimuthally symmetric gaussian model of 10\deg\ FWHM centred at 
   $l=0\deg$ and $b=0\deg$.}
   \label{fig:simulation}
\end{figure}

Figure \ref{fig:simulation} shows the resulting simulated skymap, now 
shown for the entire galactic centre region.
Obviously, although the underlying model has been azimuthally 
symmetric, the deconvolved image shows a clear emission asymmetry, 
with FWHM of $8\deg$ and $12\deg$ in the longitudinal and latitudinal 
directions, respectively.
Also the emission maximum is displaced from the galactic centre, and 
is found at $l=0.5\deg$ and $b=2.5\deg$.
These values are comparable to those found for the centroid of the 
emission in the 511~keV sky map, and as we will show more quantitatively 
in the next section, the apparent displacement in the 511~keV sky map 
could indeed be of purely statistical nature.

\section{Model fitting}
\label{sec:fitting}

A more quantitative approach is possible by fitting to the data a model of 
the spatial distribution that has one or more components.
The components can be point sources, gaussian or other geometric forms, 
or arbitrary maps corresponding to known distributions. 
The intensities of the components are adjusted to maximise the likelihood 
that the model gives rise to the observed distribution of counts in the line 
(a 5 keV wide energy band centred at 511 keV was used), binned by detector 
and by pointing.
Along with the model intensities, 19 background model scaling factors 
have been adjusted for the line component by the fit, one for each 
orbit (factors $G$ in Jean et al.~\cite{jean2003}); the scaling factors for 
the continuum component (factors $F$ in Jean et al.~\cite{jean2003}) have 
been fixed to unity.
Not fitting the background model introduces systematic uncertainties 
in the analysis that considerable biases the morphology determination.

Guided by the imaging analysis (and by previous work performed to 
describe the morphology of the OSSE observations, 
e.g. Purcell et al. \cite{purcell1997}) we used gaussian functions to 
describe the observed bulge emission.
Assuming an azimuthally symmetric gaussian centred on $l=0\deg$ and 
$b=0\deg$ results in an optimum FWHM of $9^{+7}_{-3}$ degrees, where 
the quoted uncertainties are formal statistical $2\sigma$ errors.
Systematic uncertainties, due to different treatment of the 
instrumental background, hardly affect the lower boundary, while 
adding an additional uncertainty of $\pm2\deg$ to the upper boundary.
The flux in the bulge component amounts to 
$9.9^{+4.7}_{-2.1} \times 10^{-4}$ \funit, where the uncertainty 
includes (and is dominated by) the uncertainty in the gaussian width.
Formally, the statistical detection significance of the 511 keV line 
amounts to $12\sigma$.

Relaxing the condition on spherical symmetry does not significantly 
improve the fit.
Yet relaxing the condition on the location of the bulge centroid 
improves the likelihood by $3.3$ corresponding to a significance of 
$2.1\sigma$ for the displacement.
The optimum centroid positions has been determined as to be
$l=-1.0\deg\pm1.3\deg$ and $b=1.4\deg\pm1.3\deg$,
the optimum gaussian width at this location amounts to
$8^{+4}_{-3}$ degrees ($2\sigma$ statistical errors).
Systematic uncertainties in the centroid determination amount to 
about $\pm0.2\deg$.
The flux in the displaced bulge component amounts to
$10.3^{+2.6}_{-2.2} \times 10^{-4}$ \funit, including again the 
uncertainty in the gaussian width.

Including a galactic disk component in addition to the gaussian bulge 
component does not significantly improve the fit.
In none of the considered cases did the fit attribute a significant flux 
to the disk component.
From the fits we derive upper limits on the disk flux by multiplying 
the statistical uncertainty in the disk component by the 
requested significance level (we quote here $2\sigma$ upper limits, 
hence the formal $1\sigma$ statistical errors have been multiplied by 
a factor of 2).
Although the disk models we tested formally cover the entire galactic 
plane, the effective exposure of our data is restricted to $l\approx\pm40\deg$, 
and hence we can only derive conclusions about this longitude range.
For the bulge component we used an azimuthally symmetric gaussian 
centred on $l=0\deg$ and $b=0\deg$ with the width being a free 
parameter of the fit.
In all considered cases, the best fitting FWHM amounted to 
$\sim9\deg$, with a $2\sigma$ lower limit of $6\deg$.

The resulting limits on the disk flux strongly depend on the spatial 
distribution that has been assumed for the disk component.
Using models of constant positron annihilation surface density 
throughout the galactic disk, limited to a maximum galactocentric 
radius of 14 kpc, provide the largest upper flux limits, comprised 
between $(2.8-3.4)\times 10^{-3}$ \funit\ for assumed exponential scale heights 
of 90 and 325 pc, respectively.
A tracer of the old stellar population, such as the DIRBE 35 $\mu$m allsky 
map, gives a slightly smaller limit of $2.5\times 10^{-3}$ \funit, while a 
massive star tracer, such as the DIRBE 240 $\mu$m allsky map, provides a 
considerably smaller limit of $1.4\times 10^{-3}$ \funit.

\section{Discussion}
\label{sec:discussion}

The 511 keV line emission detected by SPI from the Galaxy is so far 
adequately described by a gaussian shaped bulge of $\sim9\deg$ FWHM.
The $2\sigma$ lower limit on the bulge size amounts to $6\deg$, which 
is at the upper limit of the values suggested by the OSSE observations 
($4\deg-6\deg$).
At this early stage of the analysis we do not emphasise
this discrepancy, yet it may be taken as a first hint of morphology 
differences between the OSSE and SPI analyses.

The data do not suggest that the bulge geometry deviates from spherical 
symmetry yet a small offset of the centroid from the galactic centre direction 
is indicated by the data at the $\sim2\sigma$ confidence limit.
The present data do not yet allow to make a statement about the 
reality of the positive latitude enhancement that has been reported by 
Purcell et al. (\cite{purcell1997}).

Assuming that the bulge is indeed located at the galactic centre, at 
a distance of $8.5$ kpc, the measured 511 keV line flux converts into 
an annihilation rate of $(3.4-6.3)\times 10^{42}$ s$^{-1}$.
The observed flux is compatible with previous measurements that have 
been obtained using telescopes with small or moderate fields-of-view, 
yet it is on the low side when compared to OSSE measurements.
OSSE, however, has detected an additional galactic disk component 
that is (so far?) not seen in the SPI data.
The disk flux determined by OSSE lies in the range
$(0.7-2.7)\times 10^{-3}$ \funit, and strongly depends on the assumed shape 
of the bulge component.
Our $2\sigma$ upper limits on the disk component of 
$(1.4-3.4)\times 10^{-3}$ \funit\ 
are still compatible with the OSSE measurements.

The upper flux limits on the galactic disk component may be converted 
into lower limits for the bulge-to-disk ratio.
While the constant surface density models provide the smallest limits 
of $B/D \ge 0.3$, the DIRBE 35 $\mu$m and 240 $\mu$m maps lead to 
$B/D \ge 0.4$ and $B/D \ge 0.6$, respectively.
We included in these limits the $2\sigma$ uncertainty about the bulge 
size which translates into a typical uncertainty of $0.2$ in the 
bulge-to-disk ratio.
Again, these limits are compatible with the OSSE measurement of values 
in the $0.2-3.3$ range.

\section{Conclusions}
\label{sec:conclusions}

The analysis of SPI data that have been recorded during the first half 
of the first year's INTEGRAL GCDE have provided initial constraints on 
the morphology of the galactic 511 keV line emission.
The data suggest that the emission follows an azimuthally symmetric galactic 
bulge of $\sim9\deg$ typical FWHM, yet the uncertainties on the width 
are still rather large ($6\deg-18\deg$, $2\sigma$).
The bulge seems centred on the galactic centre, yet a marginal 
displacement towards positive latitudes and negative longitudes may be 
indicated.
Obviously, this displacement clearly needs confirmation by the 
analysis of a much larger set of data.

The available SPI observations are so far rather insensitive to a galactic 
disk component, and only upper limits have been derived.
More data along the galactic plane are needed to better constrain the 
disk component.
These data are, to some extent, already taken, yet the data 
sharing agreements do not allow for their inclusion in the present 
analysis.
Yet we are optimistic that once combined, the complete set of SPI 
observation will provide unprecedented constraints on the morphology 
of the 511 keV line emission, and thus give us key information about 
the origin of the positrons in the Galaxy.

\begin{acknowledgements}
The SPI project has been completed under the responsibility and leadership 
of CNES.
We are grateful to ASI, CEA, CNES, DLR, ESA, INTA, NASA and OSTC for 
support.
\end{acknowledgements}



\begin{thebibliography}{}

\bibitem[1997]{cheng1997} Cheng, L.~X., Leventhal, M., Smith, D.~M., 
               et al.~1997, ApJ, 481, L43

\bibitem[1998]{harris1998} Harris, M.~J., Teegarden, B.~J., Cline, T.~L., 
               et al.~1998, ApJ, 501, L55

\bibitem[2003]{jean2003} Jean, P., Kn\"odlseder, J., Lonjou, V.,
               et al.~2003, A\&A, 407, L55

\bibitem[1973]{johnson1973} Johnson, W.~N., \& Haymes, R.~C.~1973, ApJ, 
              184, 103
	      
\bibitem[1999]{knoedlseder1999} Kn\"odlseder, J., Dixon, D., Bennett, K., 
               et al.~1999, A\&A, 345, 813

\bibitem[1987]{kozlovsky1987} Kozlovsky, B., Lingenfelter, R.~E., \& 
               Ramaty, R.~1987, ApJ, 316, 801
	      
\bibitem[1978]{leventhal1978} Leventhal, M., MacCallum, C.~J., \& 
               Stang, P.~D.~1978, ApJ, 225, L11

\bibitem[1984]{lingenfelter1984} Lingenfelter, R.~E., \& Hueter, G.~J. 
               1984, in: High-Energy Transients in Astrophysics, ed. S.~E. 
               Woosley, AIP Conference Proceedings, 558	      
	      
\bibitem[1983]{lingenfelter1983} Lingenfelter, R.~E., \& Ramaty, R. 
               1983, in: Positron-Electron Pairs in Astrophysics, eds. 
               M.~L. Burns, A.~K. Harding, \& R. Ramaty, AIP Conference 
               Proceedings, 267      
	      
\bibitem[1974]{lucy1974} Lucy, L.~B.~1974, AJ, 79, 745

\bibitem[2000]{milne2000} Milne, P.~A., Kurfess, J.~D., Kinzer, R.~L.,
               Leising, M.~D., \& Dixon, D.~D.~2000, AIP Conference Proceedings, 
	       510, 21

\bibitem[2001]{milne2001} Milne, P.~A., Kurfess, J.~D., Kinzer, R.~L., 
               \& Leising, M.~D.~2001, AIP Conference 
               Proceedings, 587, 11

\bibitem[1997]{purcell1997} Purcell, W.~R., Cheng, L.-X., Dixon, 
               D.~D., et al.~1997, ApJ, 491, 725

\bibitem[1979]{ramaty1979} Ramaty, R., Kozlovsky, B., \& Lingenfelter, 
               R.~E.~1979, ApJS, 40, 487
 
\bibitem[1972]{richardson1972} Richardson, W.~H.~1972, J.~Opt.~Soc.~Am., 62, 55

\bibitem[1990]{share1990} Share, G.~H., Leising, M.~D., Messina, 
               D.~C., Purcell, W.~R.~1990, ApJ, 358, L45

\bibitem[1971]{sturrock1971} Sturrock, P.~A.~1971, ApJ, 164, 529

\bibitem[2003]{ballmoos2003} Von Ballmoos, P., Guessoum, N., Jean, 
               P., \& Kn\"odlseder, J. 2003, A\&A, 397, 635

\bibitem[2001]{winkler2001} Winkler, C. 2001,  
               Proc. 4th INTEGRAL workshop, eds. A.~Gimenez, V. Reglero \& C. 
               Winkler, (ESA SP-459) 471

\end{thebibliography}
\end{document}